\documentclass[%
 reprint,
 amsmath,amssymb,
 aps,
]{revtex4-2}

\newcommand{\nsno}{Nd$_{0.825}$Sr$_{0.175}$NiO$_2$}
\newcommand{\psno}{Pr$_{0.8}$Sr$_{0.2}$NiO$_2$}
\newcommand{\lsno}{La$_{0.8}$Sr$_{0.2}$NiO$_2$}
\newcommand{\lno}{LaNiO$_2$}
\newcommand{\sto}{SrTiO$_3$}
\newcommand{\musr}{$\mu$SR}

\usepackage{graphicx}
\usepackage{dcolumn}
\usepackage{bm}
\usepackage[numbers]{}

\begin{document}


\title{Intrinsic magnetism in superconducting infinite-layer nickelates}

\author{Jennifer Fowlie$^{1,2}$}
\email{jfowlie@stanford.edu}
\author{Marios Hadjimichael$^3$}
\author{Maria M. Martins$^{4,5}$}
\author{Danfeng Li$^{1,6}$}
\author{Motoki Osada$^{1,7}$}
\author{Bai Yang Wang$^{1,8}$}
\author{Kyuho Lee$^{1,8}$}
\author{Yonghun Lee$^{1,2}$}
\author{Zaher Salman$^4$}
\author{Thomas Prokscha$^4$}
\author{Jean-Marc Triscone$^3$}
\author{Harold Y. Hwang$^{1,2}$}
\email{hyhwang@stanford.edu}
\author{Andreas Suter$^4$}
\email{andreas.suter@psi.ch}

\affiliation{$^1$Stanford Institute for Materials and Energy Sciences, SLAC National Accelerator Laboratory, Menlo Park, CA, USA}
\affiliation{$^2$Department of Applied Physics, Stanford University, Stanford, CA, USA}
\affiliation{$^3$Department of Quantum Matter Physics, University of Geneva, Geneva, Switzerland}
\affiliation{$^4$Laboratory for Muon-Spin Spectroscopy, Paul Scherrer Institute, Villigen PSI, Switzerland}
\affiliation{$^5$Advanced Power Semiconductor Laboratory, ETH Zurich, Zurich, Switzerland}
\affiliation{$^6$Department of Physics, City University of Hong Kong, Kowloon, Hong Kong, China}
\affiliation{$^7$Department of Materials Science and Engineering, Stanford University, Stanford, CA, USA}
\affiliation{$^8$Department of Physics, Stanford University, Stanford, CA, USA}

\maketitle

The discovery of superconductivity in Nd$_{0.8}$Sr$_{0.2}$NiO$_2$ \cite{Li2019a} introduced a new family of layered nickelate superconductors that has now been extended to include a range of Sr-doping \cite{Li2020,Zeng2020}, Pr or La in place of Nd \cite{Osada2020a,Osada2021,Zeng2021}, and the 5-layer Nd$_6$Ni$_5$O$_{12}$ \cite{Pan2021}. A number of studies indicate that electron correlations are strong in these materials \cite{Botana2020,Kitatani2020a,Wu2020a,Sakakibara2020,Lechermann2020a,Werner2020,Wan2021}, and hence a central question is whether or not magnetism is present as a consequence of these interactions. Here we report muon spin rotation/relaxation studies of a series of superconducting infinite layer nickelates. In all cases we observe an intrinsic magnetic ground state, regardless of the rare earth ion or doping, arising from local moments on the nickel sublattice. The coexistence of magnetism -- which is likely to be antiferromagnetic and short-range ordered -- with superconductivity is reminiscent of some iron pnictides \cite{Stewart2011} and heavy fermion compounds \cite{Caspary1993}, and qualitatively distinct from the doped cuprates \cite{Tallon1997}.

Monovalent 3$d^9$ nickelates (Ni$^{1+}$) such as LaNiO$_2$ have long been considered in the context of divalent 3$d^9$ cuprates (Cu$^{2+}$) \cite{Rice1999,Lee2004a}. Despite being isostructural to the infinite-layer (Sr,Ca)CuO$_2$ system, it is not yet clear how similar the two families really are. The cuprates are charge-transfer insulators, and doped holes appear in the oxygen $p$-band while Cu retains its 3$d^9$ character - a situation described by the Zhang-Rice singlet \cite{Zhang1988}. In nickelates, there is a larger charge-transfer gap so the system is closer to a Mott insulator, and holes reside more predominantly in the Ni $d$-band \cite{Hepting2019a,Goodge2021,Jiang2020}. An additional aspect to consider in the nickelates is the nickel-rare earth hybridization \cite{Lee2004a}. 

A notable difference is that the parent compounds of the superconducting cuprates are long-range-ordered antiferromagnets; upon doping, dispersive magnetic excitations persist across the superconducting dome \cite{LeTacon2011}. By contrast, no long-range magnetic order has been observed in the bulk compounds LaNiO$_2$ and NdNiO$_2$ \cite{Hayward2003,Lin2021,Ortiz2021}. However, magnetic excitations consistent with the magnon dispersion expected for cuprate-like antiferromagnetic order have recently been observed in NdNiO$_2$ films \cite{Lu2021}. These features are already heavily damped; upon doping, their visibility is rapidly lost. It is therefore an open question whether magnetism is relevant for the superconducting dome in the nickelates. For this purpose, muon spin rotation/relaxation ($\mu$SR) is ideal. This is a highly local probe that also provides magnetic volume information, free from complications due to the possible role of magnetic defects or interfacial effects. Furthermore, through a comprehensive study of four infinite-layer compounds, \nsno{}, \psno{}, \lsno{}, and \lno{}, the contribution, if any, of the rare earth moment can be disentangled. 

\begin{figure*}
\includegraphics[]{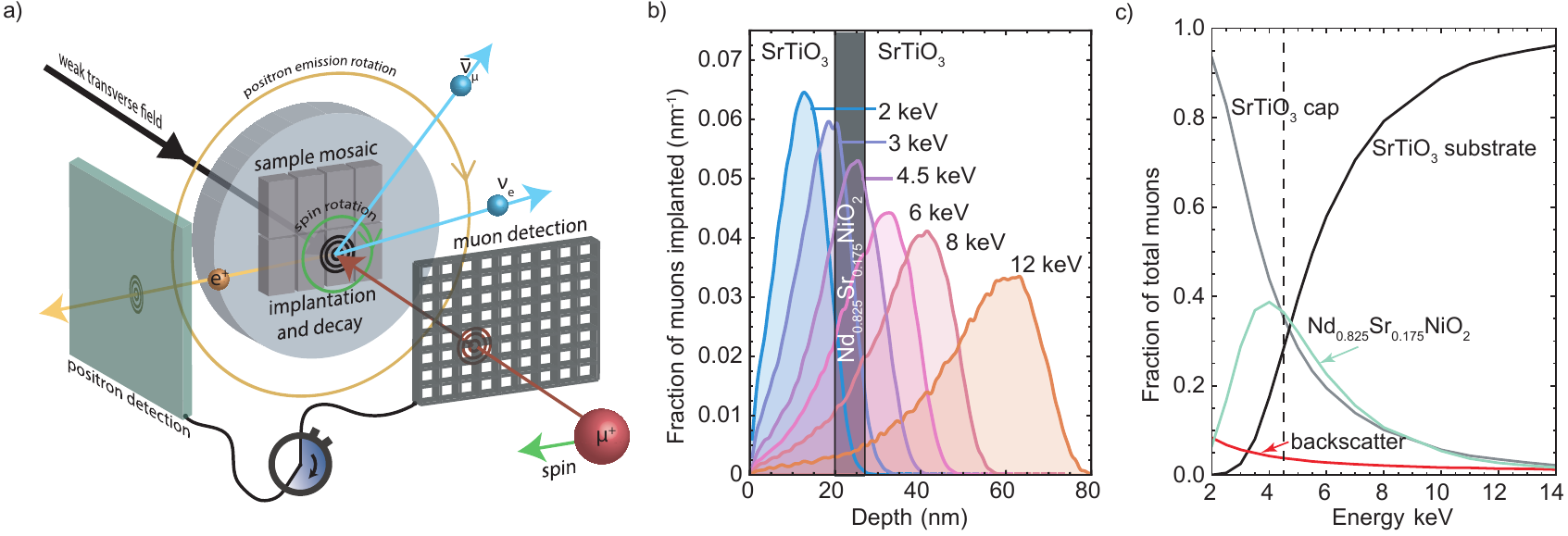}
\caption{(a), Schematic of the \musr{} set-up. The green arrow denotes the muon spin, which precesses in a transverse field (whether applied or intrinsic) and the yellow arrow illustrates the corresponding positron emission direction. Positron detectors are positioned 360$^{\circ}$ around the sample but only one is shown for simplicity. b) Simulated muon implantation profiles (a total of 200,000 muons) for the \nsno{} sample  with a 20 nm SrTiO$_3$ capping layer. c) The simulated fraction of implanted muons in \nsno{} (green), the \sto{} cap (gray), and the \sto{} substrate (black) as a function of implantation energy. Also shown is the fraction of backscattered muons in red. The dashed line at 4.5 keV is the optimal energy for the \nsno{} sample.}
\label{intro}
\end{figure*}

\musr{} involves a beam of spin-polarized, positively charged muons $\mu^+$ implanted into a sample (Fig.~\ref{intro}a). Any component of the local magnetic field $B$ that is transverse to the muon spin will cause it to precess with the Larmor frequency $\omega_{\rm L} = \gamma_\mu B$, where $\gamma_\mu$ is the gyromagnetic ratio of the muon. Muons decay by the weak force with a mean lifetime of 2.2 $\mu$s into a positron, a neutrino, and an antineutrino. The positron is emitted preferentially along the direction of the muon spin at the moment of its decay, and if the spin of the muon ensemble is precessing, so too will be the positron emission direction. Recording the positron distribution as a function of the time that the muon spent in the sample thus provides information on the local $B$. \musr{} can be used to probe the intrinsic field in the sample (``zero field'') or, alternatively, a weak magnetic field transverse to the muon spin can be applied externally; both approaches are used here. 

An important technical issue arises from the fact that high-quality crystalline superconducting infinite-layer nickelates are currently limited to thin films of $\sim$10 nm thickness, and often require a capping layer for stability (\sto{} is both the substrate and capping layer in these experiments). By decreasing the muon beam energy down to a few keV, the stopping depth is reduced to the required nanometer scale. However, below 2 keV a significant fraction of the muons are lost due to backscattering and reflection. Thus we used the capping layer thickness and Monte Carlo simulations to design the optimal heterostructures and beam energies for 8 nm thick nickelate layers (Supplementary Information). Fig.~\ref{intro}b,c shows representative calculated muon stopping profiles for the \nsno{} sample. 

\begin{figure*}
\includegraphics[]{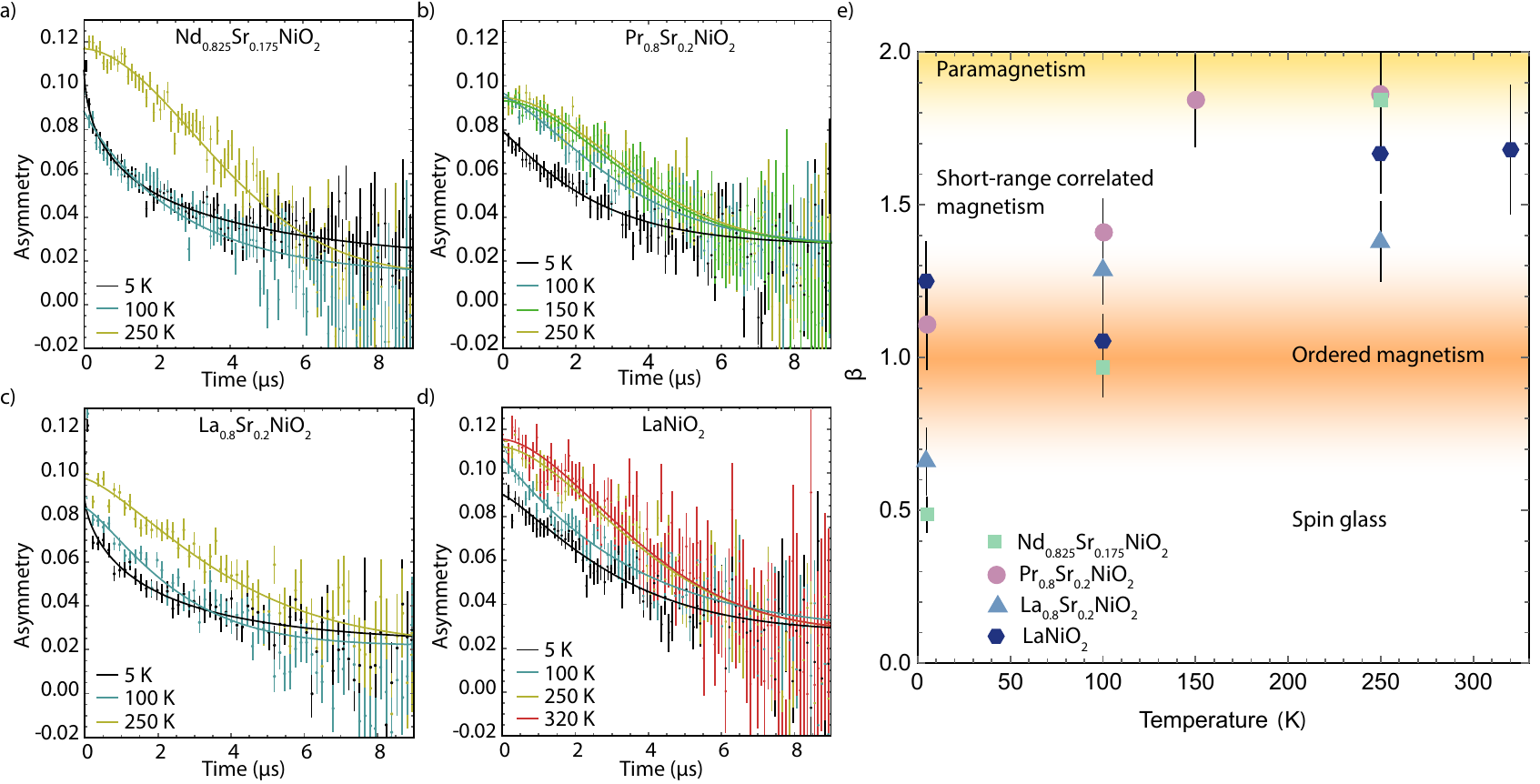}
\caption{(a-d) Asymmetry in zero field at various temperatures for \nsno{}, \psno{}, \lsno{}, and \lno{}. Error bars represent one standard deviation of a Poisson distribution for each time bin. e) The $\beta$ parameter from the stretched exponential fits as a function of temperature for the four samples. Colored regions indicate $\beta$ values for different forms of magnetism. Error bars represent one standard deviation from the fit.}
\label{ZF}
\end{figure*}

Using this approach, we measured the zero field (ZF) asymmetry of the four infinite-layer nickelate compositions at various temperatures down to 5 K (Fig.~\ref{ZF}a-d). A clear temperature dependence is observed, indicating the appearance of local magnetism. No ZF precession is observed, suggesting dephasing due to a broad field distribution and/or relaxation due to fluctuations.
To analyze these data, we consider that in a magnetic system with a \emph{randomly-oriented} Maxwell-Boltzmann distribution of local magnetic fields, the ZF muon depolarization along the direction of initial spin is described by a Gaussian-Kubo-Toyabe (GKT) function \cite{Kubo1967}:
\begin{equation}
P^{\rm GKT}(t)=\frac{1}{3}+\frac{2}{3}(1-\sigma^2t^2)\cdot \exp(-\frac{\sigma^2t^2}{2}).
\label{GKT}
\end{equation}
Here $t$ is the time in $\mu$s and $\sigma$ is the depolarization rate in $\mu$s$^{-1}$. The $\frac{1}{3}$ and $\frac{2}{3}$ coefficients are relaxed to additional fit parameters $A_{\rm BG}$ and $A_0$ -- background and initial asymmetry, respectively -- because the local field is not expected to be randomly oriented and distributed in this thin film system. Furthermore, not all muons stop in the nickelate layer. With this in mind, and allowing for other forms of magnetism, we generalize to a stretched exponential function of the form:
\begin{equation}
A(t)=A_0\, \exp(-\lambda t^\beta)+A_{\rm BG}.
\label{strexp}
\end{equation} 

In a paramagnetic system, electronic moments fluctuate rapidly and do not contribute to the overall depolarization, leaving only very weak nuclear moments to depolarize the muon ensemble. In this limit (small $\sigma$), the GKT function reduces to a Gaussian function and is equivalent to Equation \ref{strexp} with $\lambda=\frac{1}{2}\sigma^2$ and $\beta$ = 2. In a magnetically-ordered system with strong and rapidly-fluctuating local moments, an exponentially decaying depolarization function is expected \cite{Garcia-Munoz1995,Chow1996}, thus $\beta$ = 1. The Gaussian to exponential crossover is well-documented in similar bulk materials undergoing independently-verifiable magnetic ordering, such as the stripe-ordered Ruddlesden-Popper nickelates \cite{Chow1996} and perovskite nickelates \cite{Garcia-Munoz1995}. Finally, a stretched exponential depolarization function with $\beta \leq \frac{1}{2}$ is found for spin glass systems \cite{Campbell1994,Keren1996}. Therefore, $\beta$ is a useful metric characterizing the presence and nature of magnetism.

The solid lines in Fig.~\ref{ZF}a-d give stretched exponential fits to our data, and the resulting $\beta$ parameters for all samples and measurement temperatures are plotted in Fig.~\ref{ZF}e (see Supplementary Information for full parameters). All four samples approach the paramagnetic limit ($\beta$ = 2) at high temperatures, with some indications of residual short-range magnetic correlations in \lno{} and particularly \lsno{}. $\beta$ decreases upon cooling down to 5 K, where \lno{} and \psno{} have $\beta$ parameters $\sim$1 indicative of magnetic ordering, while \nsno{} and \lsno{} appear to approach the regime of spin glasses.

ZF-\musr, therefore, shows that all the samples exhibit some form of intrinsic low-temperature magnetism. For \nsno{}, \psno{}, and \lsno{}, the 5 K measurement is within the superconducting state (Supplementary Information) so this provides direct evidence for coexisting superconductivity and magnetism in infinite-layer nickelates. To determine if this coexistence is relevant down to the microscopic level, weak transverse field (wTF) \musr{} measurements were performed to obtain the magnetic volume fraction of these samples.

The muon acts as a local magnetic sensor and there are two extremal cases under an applied transverse magnetic field: 1. If the internal field is much smaller than the applied field, the muon spin ensemble experiences dominantly the applied field. In this case the muon spin precession is around the applied field and has a maximal amplitude (asymmetry). 2. If the local magnetic field distribution at the muon site is comparable to or stronger than the applied field, the muon spin ensemble dephases very rapidly. Therefore, the amplitude of the asymmetry of the resulting muon precession is proportional to the non-magnetic volume fraction of the sample. 
Each measurement, from 5 K up to room temperature, consists of at least two million events and is performed in a 10 mT field applied transverse to the initial muon spin polarization (see Fig.~\ref{intro}a). 

\begin{figure*}
\includegraphics[]{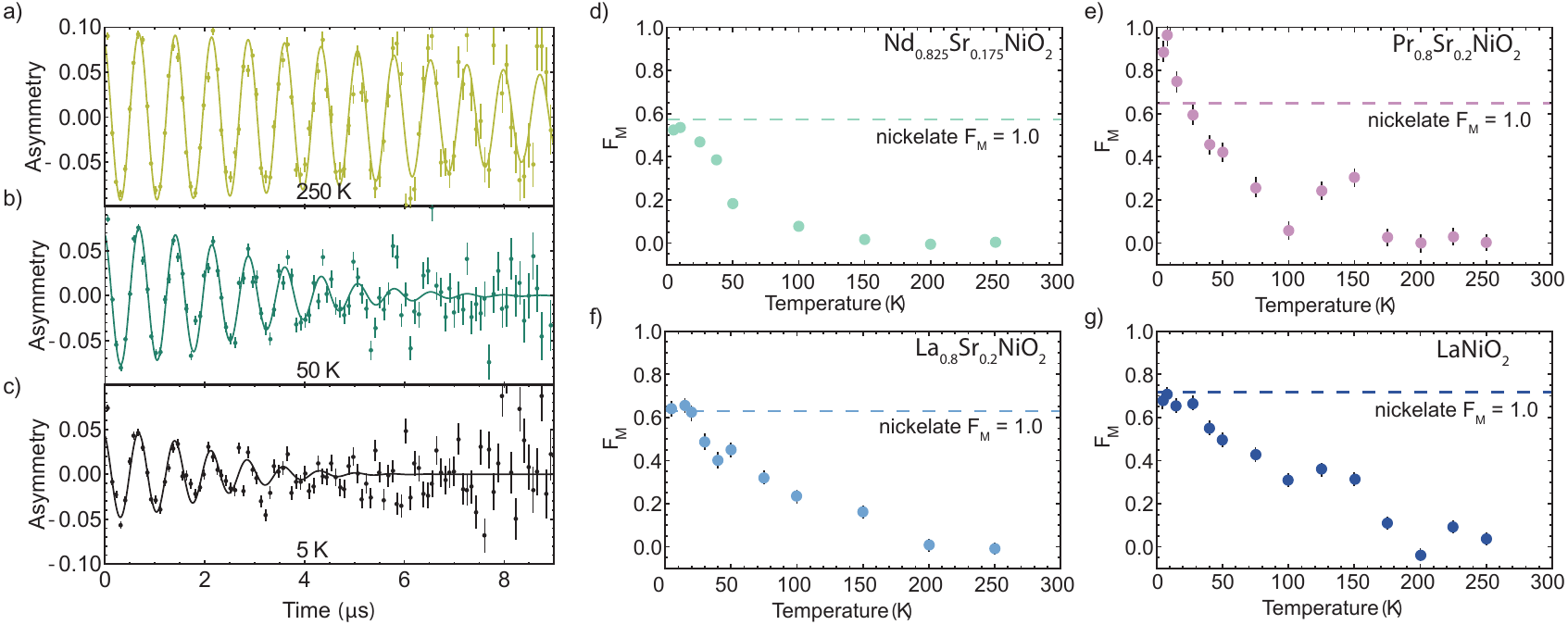}
\caption{(a-c) Asymmetry in a weak transverse field of $B$ = 10 mT for the \nsno{} sample at 250 K (a), 50 K (b), and 5 K (c). The solid line is a decaying cosine from which the initial asymmetry amplitude is extracted. Error bars represent one standard deviation of a Poisson distribution for each time bin. (d-g), Magnetic volume fraction ($F_{\rm M}$) as a function of temperature for the four samples, \nsno{}, \psno{}, \lsno{}, and \lno{}. The dashed lines correspond to the expected $F_{\rm M}$ when 100\% of the nickelate is magnetic. Error bars represent one standard deviation from the fit.}
\label{Tscan}
\end{figure*}

Fig.~\ref{Tscan}a-c displays the oscillations arising from the muon precession around the transverse field for \nsno{} at 250, 50, and 5 K. The solid lines represent a decaying cosine fit:
\begin{equation}
A(t)=A_0 e^{-\lambda_{\rm TF} t} \cos(\gamma_{\mu}Bt+\phi).
\end{equation}
Here, $\phi$ is the relative detector phase with respect to the initial muon spin orientation. With decreasing temperature, the decreasing initial asymmetry $A_0$, as well as the increasing depolarization rate $\lambda_{\rm TF}$, both indicate the onset of magnetism at low temperatures. From $A_0$ the magnetic volume fraction, $F_{\rm M}$, can be calculated as:
\begin{equation}
F_{\rm M}(T)=1-\frac{A_0(T)}{A_{\rm PM}}.
\end{equation}
To evaluate $F_{\rm M}$ it is assumed that the high temperature saturation of $A_0$ corresponds to a paramagnetic state and only nuclear moments remain. The paramagnetic asymmetry $A_{\rm PM}$ is estimated as the mean of the initial asymmetries above 200 K.
Fig.~\ref{Tscan}d-g plots $F_{\rm M}$ as a function of temperature for all samples. Since not all of the muons are stopped in the nickelate layer, the dashed lines represent a 100\% magnetic nickelate. All four samples exhibit an increasing $F_{\rm M}$ with decreasing temperature, reaching 100\% magnetic volume at the lowest temperatures. This agrees well with the picture suggested by the ZF measurements that the superconducting infinite layer nickelates are intrinsically magnetic, and further reinforces the conclusion that superconductivity and magnetism coexist.

\begin{figure}
\includegraphics[]{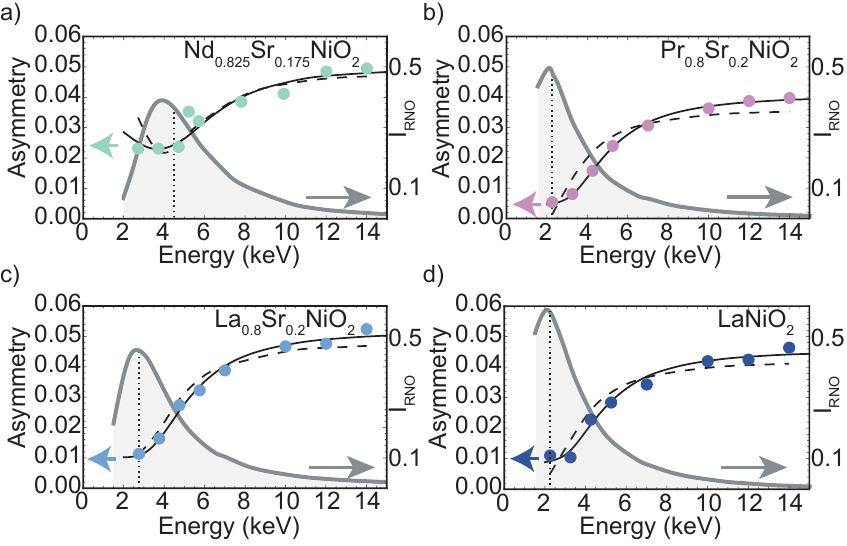}
\caption{wTF muon decay asymmetry at 15 K as a function of energy for \nsno{} (a), \psno{} (b), \lsno{} (c), and \lno{} (d). The right axis is the fraction of the total implanted muons that are stopped in the nickelate layer as determined from Monte Carlo simulations. The dashed and solid lines are fits to the experimental data assuming sharp changes of asymmetry at the interfaces in the multilayer (dashed) and allowing the asymmetry to ``leak'' across the interfaces (solid) as described in the text. The vertical dotted lines represent the central energies that are used for the temperature-dependent measurements in Figs.~\ref{ZF} and \ref{Tscan}. Uncertainties representing one standard deviation from the fit are smaller than the plot markers.}
\label{Escan}
\end{figure}

Finally, the implantation energy can be varied as a proxy for a depth scan. This was carried out for all four samples at 15 K (Fig.~\ref{Escan}); each panel shows the initial asymmetry (left axis) as a function of energy, as well as the simulated fraction of muons (right axis) implanted in the nickelate layer for a given energy.  For all four samples the energy-dependent trend is similar, with a minimal asymmetry coinciding with the maximal muon implantation into the nickelate layer. Furthermore, when $E \gtrsim 10\,$keV the asymmetry begins to saturate at around 0.04-0.05. This behavior can be fitted with a simple model based on the calculated implantation profiles and assuming step-function changes in the asymmetry at the interfaces in the heterostructure (dashed lines in Fig.~\ref{Escan}). When the asymmetry of the nickelate layer is allowed to penetrate into the substrate, i.e., representing demagnetizing fields, the fit result is given by the solid lines. These minimal models agree well with each other and with the experimental data, which suggests that the magnetic properties change sharply close to the interfaces between the nickelate layer and the \sto{} on either side. 
This indicates that long-range demagnetizing fields, as would be expected from a ferromagnetic thin film, are minimal in these nickelates. The energy-dependence, therefore, suggests that the intrinsic magnetism in the infinite layer nickelates is based on an antiferromagnetic coupling.

This depth-dependent information, together with the temperature-dependence of the wTF and ZF spectra, offers clear evidence for intrinsic magnetism in the infinite-layer nickelates. We now briefly discuss the implications of our work.

The undoped \lno{} sample is magnetic with approximately 100\% volume fraction at low temperature despite the absence of a magnetic moment on the La$^{3+}$ ion. This observation, combined with the close similarity of both the ZF and wTF results upon exchanging the rare earth in the doped materials, indicates that the observed magnetism originates from the nickel sublattice.
In going from \lno{} to \lsno{}, the magnetic state evolves to a spin-glass-like behavior -- while maintaining the full volume at low temperatures. This is consistent with spin-dilution and solid state disorder brought about by hole doping with Sr, and may provide an explanation for the doping evolution of magnetic excitations in the Nd-based compounds \cite{Lu2021}.

The temperature-dependence of $F_{\rm M}$ shows that the magnetism onsets gradually. This, together with the lack of ZF precession, suggests that the magnetic state is not static long-range ordered and that fluctuations (faster than the muon decay timescale) may be present and short-range correlations may dominate. This is consistent with previous reports of a lack of long-range order in the undoped infinite-layer nickelates \cite{Hayward2003,Lin2021,Ortiz2021}.

The short-range-ordered and glassy behavior that we report here is reminiscent of some cuprate superconductors where the spin glass state, with increased hole doping, persists into the superconducting dome \cite{Tallon1997,Stilp2013}. However, a striking difference in this context is that the spin glass in cuprates usually appears at $<$ 30 K, much lower than observed here in nickelates where an evolution begins already at $\sim$150 K. This is despite the larger superexchange energies of cuprates ($100 \lesssim J \lesssim 150$ meV) \cite{Bourges1997,LeTacon2011} versus nickelates ($50 \lesssim J \lesssim 100$ meV) \cite{Lu2021,Lin2021a}.  This high onset temperature is more similar to the antiferromagnetic transition temperature of many iron pnictides where magnetic order and superconductivity coexist \cite{Stewart2011}.

Our work demonstrates that the family of infinite-layer nickelates is intrinsically magnetic, including in the superconducting state. This coexistence, together with the high temperature onset, highlights a distinction from the nominally-similar cuprates and suggests a more complex picture where the multi-orbital nature of the nickelates may play an important role.

\section*{Methods}
The nickelate thin films capped with \sto{} were deposited on \sto{} substrate by pulsed laser deposition as the respective perovskite phase before being topotactically reduced by a soft chemical process as described in Section A of the Supplementary Information and elsewhere \cite{Lee2020}. 
Low energy $\mu$SR was carried out at the LEM facility at the $\mu$E4 beamline of the Swiss Muon Source at PSI ~\cite{Prokscha2008}. The sample mosaic was mounted on a nickel coated plate in order to have a flat background asymmetry \cite{Saadaoui2012}. All the \musr{} data have been analyzed using musrfit \cite{Suter2012}. The ZF and wTF spectra were fit from 0.07 - 9 $\mu$s. In order to estimate the effective magnetic volume fraction of fully magnetic nickelate layers (dashed lines in Fig.~\ref{Tscan}d-g.), reduced asymmetry due to muonium formation in the \sto{} has been accounted for \cite{Salman2014}.
The energy-dependent models were generated from a MATLAB routine \cite{Simoes2020}.

\begin{acknowledgments}
The work at Stanford/SLAC was supported by the US Department of Energy, Office of Basic Energy Sciences, Division of Materials Sciences and Engineering, under contract number DE-AC02-76SF00515, and Gordon and Betty Moore Foundation’s Emergent Phenomena in Quantum Systems Initiative through grant number GBMF9072 (synthesis equipment). J.F. was also supported by the Swiss National Science Foundation through Postdoc.Mobility P400P2199297 and Division II 200020\_179155. J.F., M.H. and J.M.T. acknowledge support from the Swiss National Science Foundation through Division II 200020\_179155 and the European Research Council under the European Union's Seventh Framework Program (FP7/2007-2013)/ERC Grant Agreement No.~319286 (Q-MAC). D.L. acknowledges support from Hong Kong Research Grant Council (CityU 21301221) and National Natural Science Foundation of China (12174325).
Part of this work is based on experiments performed at the Swiss Muon Source S$\mu$S, Paul Scherrer Institute, Villigen, Switzerland.
\end{acknowledgments}

\bibliography{PAPERS-MuSR}

\end{document}


\title{Supplementary information for ``intrinsic magnetism in superconducting infinite-layer nickelates''}
\author{Jennifer Fowlie$^{1,2}$}
\email{jfowlie@stanford.edu}
\author{Marios Hadjimichael$^3$}
\author{Maria M. Martins$^{4,5}$}
\author{Danfeng Li$^{1,6}$}
\author{Motoki Osada$^{1,7}$}
\author{Bai Yang Wang$^{1,8}$}
\author{Kyuho Lee$^{1,8}$}
\author{Yonghun Lee$^{1,2}$}
\author{Zaher Salman$^4$}
\author{Thomas Prokscha$^4$}
\author{Jean-Marc Triscone$^3$}
\author{Harold Y. Hwang$^{1,2}$}
\email{hyhwang@stanford.edu}
\author{Andreas Suter$^4$}
\email{andreas.suter@psi.ch}

\affiliation{$^1$Stanford Institute for Materials and Energy Sciences, SLAC National Accelerator Laboratory, Menlo Park, CA, USA}
\affiliation{$^2$Department of Applied Physics, Stanford University, Stanford, CA, USA}
\affiliation{$^3$Department of Quantum Matter Physics, University of Geneva, Geneva, Switzerland}
\affiliation{$^4$Laboratory for Muon-Spin Spectroscopy, Paul Scherrer Institute, Villigen PSI, Switzerland}
\affiliation{$^5$Advanced Power Semiconductor Laboratory, ETH Zurich, Zurich, Switzerland}
\affiliation{$^6$Department of Physics, City University of Hong Kong, Kowloon, Hong Kong, China}
\affiliation{$^7$Department of Materials Science and Engineering, Stanford University, Stanford, CA, USA}
\affiliation{$^8$Department of Physics, Stanford University, Stanford, CA, USA}
\maketitle

\subsection{\label{sec:level1}Sample preparation and characterization}

The perovskite precursor samples are grown by pulsed laser deposition on single crystal \sto{} substrates. The substrates are annealed prior to growth for 30-70 minutes at 930 $^{\circ}$C in an oxygen partial pressure of 5 $\times$ 10$^{-6}$ Torr. The nickelate film and \sto{} capping layer are grown at 570 $^{\circ}$C in an oxygen partial pressure of 0.2-0.25 Torr from polycrystalline pressed powder targets. The excimer laser has a spot fluence of 2 Jcm$^{-2}$ and is operated at 4 Hz. \nsnop{} is grown on one 10 $\times$ 10 mm substrate while \psnop{}, \lsnop{} and \lnop{} are synthesized in 3-4 growths on substrates with 5 $\times$ 5 mm dimensions.

The perovskite films are then cut in half and reduced to the infinite-layer phase. For this, they are embedded in 0.1 g of CaH$_2$ powder and heated to 240 - 260 $^{\circ}$C typically multiple times until the correct phase is observed by x-ray diffraction (XRD). The total reduction time varies from 2 - 12 hours.

\begin{figure}
\includegraphics[]{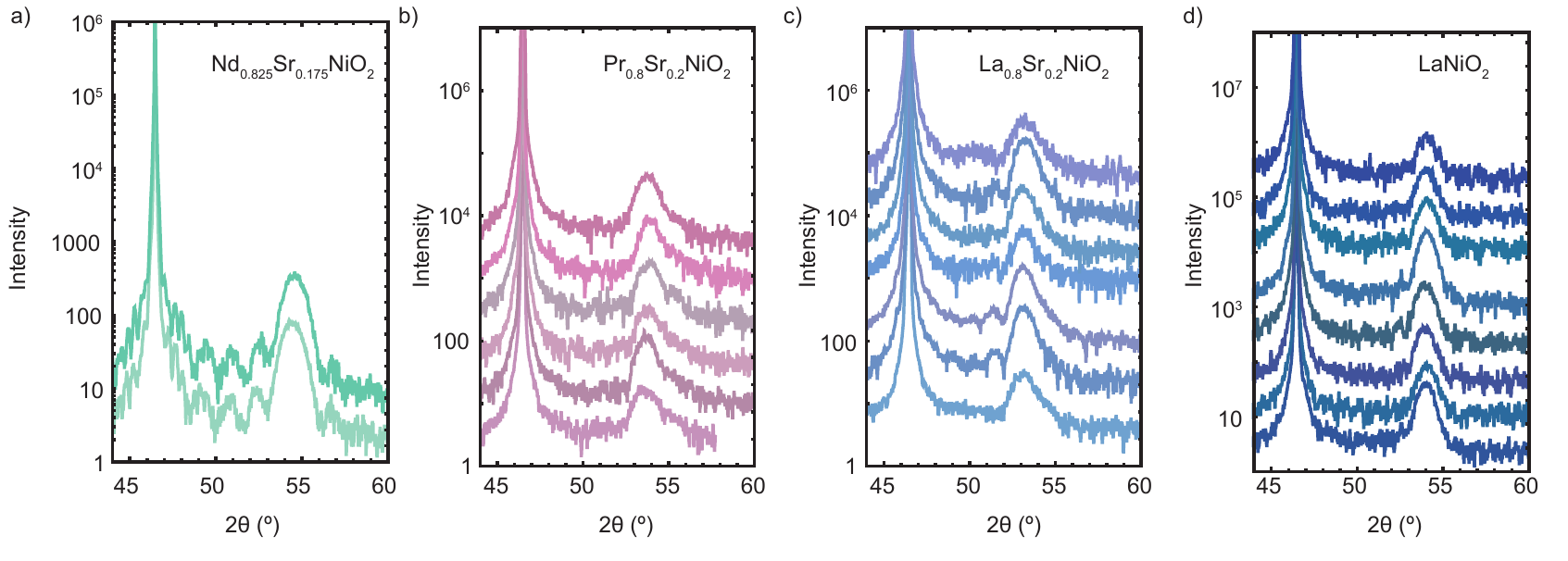}
\caption{X-ray diffraction of the infinite-layer films. a)-d) $\theta$-2$\theta$ scans around the 002 diffraction peak of each sample from the mosaic of \nsno{}, \psno{}, \lsno{} and \lno{} respectively.}
\label{XRD}
\end{figure}

Figure \ref{XRD} shows the XRD normal $\theta$-2$\theta$ scans around the 002 diffraction peak after the final reduction of each piece of the mosaic for each sample.

\begin{figure}
\includegraphics[]{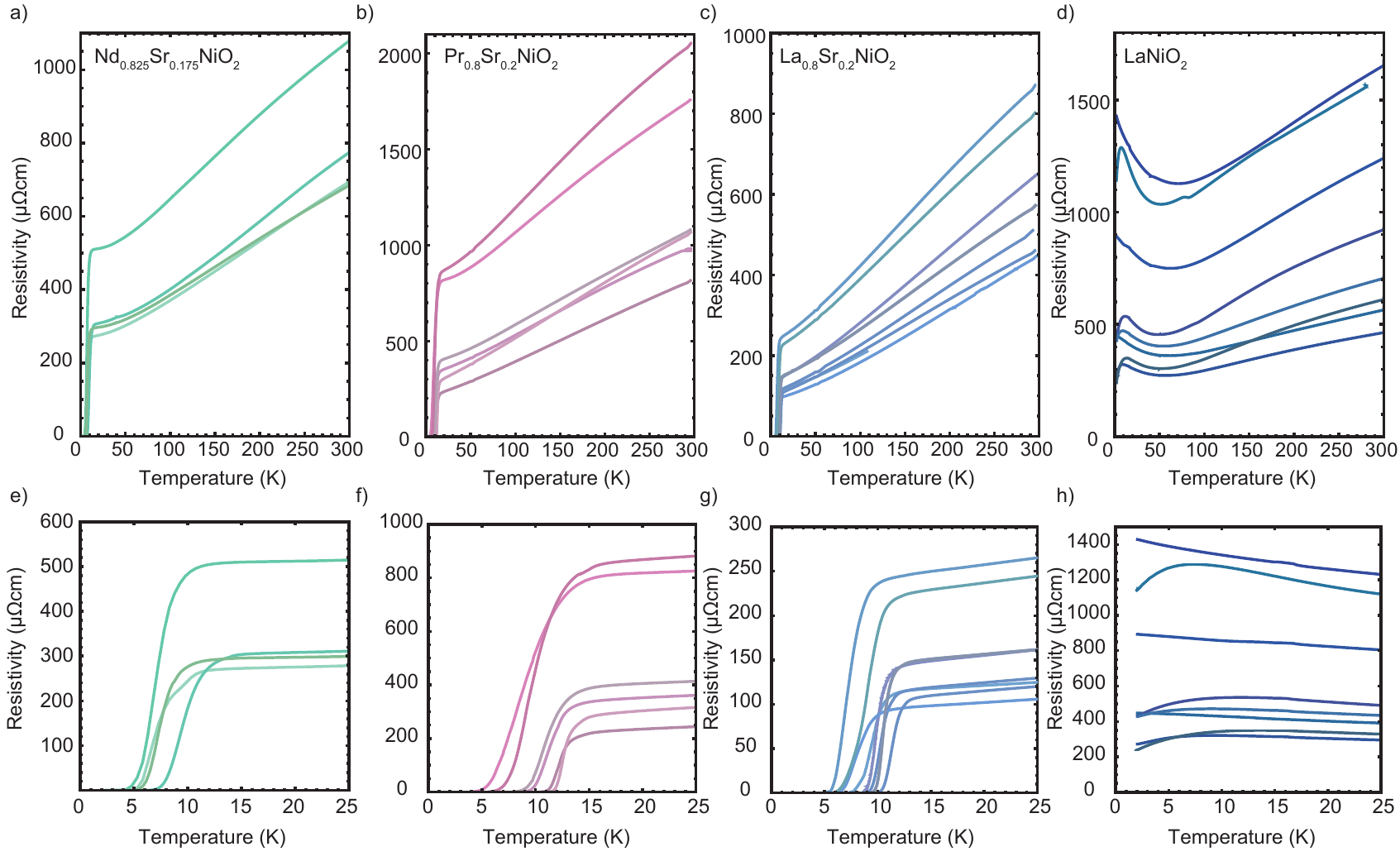}
\caption{Transport properties. a)-d) Resistivity up to 300 K for all the films. e)-h) Low temperature resistivity behavior for the same films.}
\label{RT}
\end{figure}

Figure \ref{RT} shows the resistivity as a function of temperature for all the pieces of each sample mosaic (note that the \nsno{} mosaic consists of two pieces only but transport measurements were made at two different locations on each sample). Panels a)-d) are over the full temperature range measured while panels e)-h) focus on the low temperature behavior. The superconducting transitions for the Sr-doped samples are clearly observed between 5 and 15 K while \lno{} exhibits either weakly insulating behavior or a resistivity downturn below 10 K suggesting proximity to a superconducting ground state.

\FloatBarrier
Table \ref{sampleproperties} summarizes the mean thicknesses and critical temperatures of the sample mosaics. The nickelate thicknesses are obtained from the XRD $\theta$-2$\theta$ peak widths via the Scherrer approximation \cite{Lee2020}. T$_c$ is defined as the temperature of maximum derivative of the resistivity.
\begin{table}[h]
\centering
\begin{tabular}{| c | c | c | c | c |} 
 \hline
 Sample & Nickelate thickness (nm) & Nickelate c-axis lattice parameter (\r{A}) & \sto{} cap thickness (nm) & T$_c$ (K) \\ [0.5ex]
 \hline
 \nsno{} & 7.0 & 3.363 & 20.0 & 8.4 \\ 
 \hline
 \psno{} & 7.7 & 3.405 & 9.6 & 10.5 \\ 
 \hline
 \lsno{} & 7.7 & 3.443 & 13.6 & 9.1 \\ 
 \hline
 \lno{} & 9.1 & 3.391 & 8.9 & - \\ 
 [1ex] 
 \hline
\end{tabular}
\caption{Mean sample properties}
\label{sampleproperties}
\end{table}

To calculate the optimal implantation energy, Monte Carlo simulations were performed using the TRIM.SP code \cite{Eckstein1991}. The energies found to maximize the implantation in the nickelate layers strongly depend on the thickness of the \sto{} capping layer and the density of the nickelate layer and so have to be determined by separately inputting the nominal densities and measured thicknesses of each sample. They are found to be 4.5 keV, for \nsno{}, 2.25 keV for \psno{}, 2.75 keV for \lsno{} and 2.25 keV for \lno{}.

\FloatBarrier
\subsection{Run log}
Tables \ref{runlogzf}, \ref{runlogwtfT} and \ref{runlogwtfE} contain a comprehensive list of all the low-energy muon (LEM) measurements performed in zero field, in a weak transverse field (wTF) as a function of temperature and in wTF as a function of energy respectively. For clarity the measurements are also indexed with the conditions. The run number and year can be used to call the asymmetry histogram files on musruser.psi.ch.

\begin{table}[h]
\centering
\begin{tabular}{| c | c | c | c | c | c |} 
 \hline
 Sample & T (K) & E (keV) & B (mT) & Run no. & Year \\ [0.5ex]
 \hline
 \nsno{} & 5 & 4.5 & 0 & 3822 & 2020 \\ 
 & 100 & 4.5 & 0 & 4782 & 2021 \\
 & 250 & 4.5 & 0 & 3828 & 2020 \\
 \hline
 \psno{} & 5 & 2.25 & 0 & 4750 & 2021 \\ 
 & 100 & 2.25 & 0 & 4771 & 2021 \\
 & 150 & 2.25 & 0 & 4772 & 2021 \\
 & 250 & 2.25 & 0 & 4770 & 2021 \\
 \hline
 \lsno{} & 4.7 & 2.75 & 0 & 4708 & 2021 \\ 
 & 100 & 2.75 & 0 & 4785 & 2021 \\
 & 250 & 2.75 & 0 & 4722 & 2021 \\
 \hline
 \lno{} & 4.5 & 2.25 & 0 & 4727 & 2021 \\ 
 & 100 & 2.25 & 0 & 4787 & 2021 \\
 & 250 & 2.25 & 0 & 4747 & 2021 \\
 & 320 & 2.25 & 0 & 4789 & 2021 \\
 [1ex] 
 \hline
\end{tabular}
\caption{\musr{} run log index for zero field asymmetry histograms.}
\label{runlogzf}
\end{table}

\begin{table}[]
\centering
\begin{tabular}{| c | c | c | c | c | c |} 
 \hline
 Sample & T (K) & E (keV) & B (mT) & Run no. & Year \\ [0.5ex]
 \hline
 \nsno{} & 5 & 4.5 & 10 & 3812 & 2020 \\ 
 & 10 & 4.5 & 10 & 3811 & 2020 \\
 & 25 & 4.5 & 10 & 3810 & 2020 \\
 & 37.5 & 4.5 & 10 & 3827 & 2020 \\
 & 50 & 4.5 & 10 & 3809 & 2020 \\
 & 100 & 4.5 & 10 & 3808 & 2020 \\
 & 150 & 4.5 & 10 & 3807 & 2020 \\
 & 200 & 4.5 & 10 & 3806 & 2020 \\
 & 250 & 4.5 & 10 & 3829 & 2020 \\
 \hline
 \psno{} & 5 & 2.25 & 10 & 4749 & 2021 \\ 
 & 8 & 2.25 & 10 & 4751 & 2021 \\
 & 15 & 2.25 & 10 & 4752 & 2021 \\
 & 27.5 & 2.25 & 10 & 4760 & 2021 \\
 & 40 & 2.25 & 10 & 4761 & 2021 \\
 & 50 & 2.25 & 10 & 4762 & 2021 \\
 & 75 & 2.25 & 10 & 4763 & 2021 \\
 & 100 & 2.25 & 10 & 4764 & 2021 \\
 & 125 & 2.25 & 10 & 4765 & 2021 \\
 & 150 & 2.25 & 10 & 4767 & 2021 \\
 & 175 & 2.25 & 10 & 4768 & 2021 \\
 & 200 & 2.25 & 10 & 4769 & 2021 \\
 & 225 & 2.25 & 10 & 4768 & 2021 \\
 & 250 & 2.25 & 10 & 4769 & 2021 \\
 \hline
 \lsno{} & 4.7 & 2.75 & 10 & 4707 & 2021 \\ 
 & 15 & 2.75 & 10 & 4709 & 2021 \\
 & 20 & 2.75 & 10 & 4706 & 2021 \\
 & 30 & 2.75 & 10 & 4715 & 2021 \\
 & 40 & 2.75 & 10 & 4716 & 2021 \\
 & 50 & 2.75 & 10 & 4717 & 2021 \\
 & 75 & 2.75 & 10 & 4718 & 2021 \\
 & 100 & 2.75 & 10 & 4719 & 2021 \\
 & 150 & 2.75 & 10 & 4720 & 2021 \\
 & 200 & 2.75 & 10 & 4705 & 2021 \\
 & 250 & 2.75 & 10 & 4721 & 2021 \\
 \hline
 \lno{} & 4.5 & 2.25 & 10 & 4726 & 2021 \\
 & 8 & 2.25 & 10 & 4728 & 2021 \\
 & 15 & 2.25 & 10 & 4729 & 2021 \\
 & 27.5 & 2.25 & 10 & 4737 & 2021 \\
 & 40 & 2.25 & 10 & 4738 & 2021 \\
 & 50 & 2.25 & 10 & 4739 & 2021 \\
 & 75 & 2.25 & 10 & 4740 & 2021 \\
 & 100 & 2.25 & 10 & 4741 & 2021 \\
 & 125 & 2.25 & 10 & 4742 & 2021 \\
 & 150 & 2.25 & 10 & 4743 & 2021 \\
 & 175 & 2.25 & 10 & 4744 & 2021 \\
 & 200 & 2.25 & 10 & 4725 & 2021 \\
 & 225 & 2.25 & 10 & 4745 & 2021 \\
 & 250 & 2.25 & 10 & 4746 & 2021 \\
 [1ex] 
 \hline
\end{tabular}
\caption{\musr{} run log index for weak transverse field asymmetry histograms at various temperatures.}
\label{runlogwtfT}
\end{table}

\begin{table}[h]
\centering
\begin{tabular}{| c | c | c | c | c | c |} 
 \hline
 Sample & T (K) & E (keV) & B (mT) & Run no. & Year \\ [0.5ex] 
 \hline
 \nsno{} & 15 & 2.7 & 10 & 4773 & 2021 \\
 & 15 & 3.7 & 10 & 4774 & 2021 \\ 
 & 15 & 4.7 & 10 & 4776 & 2021 \\
 & 15 & 5.2 & 10 & 4775 & 2021 \\ 
 & 15 & 5.7 & 10 & 4777 & 2021 \\ 
 & 15 & 7.8 & 10 & 4778 & 2021 \\
 & 15 & 9.9 & 10 & 4779 & 2021 \\ 
 & 15 & 12 & 10 & 4780 & 2021 \\ 
 & 15 & 14 & 10 & 4781 & 2021 \\ 
 \hline
 \psno{} & 15 & 2.25 & 10 & 4752 & 2021 \\
 & 15 & 3.25 & 10 & 4753 & 2021 \\
 & 15 & 4.25 & 10 & 4754 & 2021 \\
 & 15 & 5.25 & 10 & 4755 & 2021 \\
 & 15 & 7 & 10 & 4756 & 2021 \\
 & 15 & 10 & 10 & 4757 & 2021 \\
 & 15 & 12 & 10 & 4758 & 2021 \\
 & 15 & 14 & 10 & 4759 & 2021 \\
 \hline
 \lsno{} & 15 & 2.75 & 10 & 4709 & 2021 \\
 & 15 & 3.75 & 10 & 4710 & 2021 \\
 & 15 & 4.75 & 10 & 4711 & 2021 \\
 & 15 & 5.75 & 10 & 4723 & 2021 \\
 & 15 & 7 & 10 & 4712 & 2021 \\
 & 15 & 10 & 10 & 4713 & 2021 \\
 & 15 & 12 & 10 & 4724 & 2021 \\
 & 15 & 14 & 10 & 4714 & 2021 \\
 \hline
 \lno{} & 15 & 2.25 & 10 & 4729 & 2021 \\
 & 15 & 3.25 & 10 & 4730 & 2021 \\
 & 15 & 4.25 & 10 & 4731 & 2021 \\
 & 15 & 5.25 & 10 & 4732 & 2021 \\
 & 15 & 7 & 10 & 4733 & 2021 \\
 & 15 & 10 & 10 & 4734 & 2021 \\
 & 15 & 12 & 10 & 4735 & 2021 \\
 & 15 & 14 & 10 & 4736 & 2021 \\
 [1ex] 
 \hline
\end{tabular}
\caption{\musr{} run log index for weak transverse field asymmetry histograms at various implantation energies.}
\label{runlogwtfE}
\end{table}

\FloatBarrier
\subsection{\musr{} fit parameters}

\begin{table}[h]
\centering
\begin{tabular}{| c | c | c | c | c | c |} 
 \hline
 Sample & T (K) & $A_0$ & $\lambda$ ($\mu s^{-1}$) & $\beta$ & $A_{\rm BG}$ \\ [0.5ex] 
 \hline
 \nsno{} & 5 & 0.1011 $\pm$ 0.0091 & 0.495 $\pm$ 0.112 & 0.487 $\pm$ 0.057 & 0.0132 \\ 
 & 100 & 0.0755 $\pm$ 0.0035 & 0.374 $\pm$ 0.025 & 0.969 $\pm$ 0.097 & 0.0132 \\
 & 250 & 0.1037 $\pm$ 0.0012 & 0.2201 $\pm$ 0.0039 & 1.841 $\pm$ 0.095 & 0.0132 \\
 \hline
 \psno{} & 5 & 0.0518 $\pm$ 0.0031 & 0.384 $\pm$ 0.031 & 1.11 $\pm$ 0.16 & 0.0272 \\ 
 & 100 & 0.0675 $\pm$ 0.0016 & 0.2818 $\pm$ 0.0094 & 1.41 $\pm$ 0.11 & 0.0272 \\
 & 150 & 0.0658 $\pm$ 0.0013 & 0.2408 $\pm$ 0.0068 & 1.84 $\pm$ 0.16 & 0.0272 \\
 & 250 & 0.0672 $\pm$ 0.0012 & 0.2343 $\pm$ 0.0065 & 1.86 $\pm$ 0.15 & 0.0272 \\
 \hline
 \lsno{} & 4.7 & 0.0683 $\pm$ 0.0085 & 0.517 $\pm$ 0.129 & 0.66 $\pm$ 0.11 & 0.0215 \\ 
 & 100 & 0.0632 $\pm$ 0.0021 & 0.374 $\pm$ 0.017 & 1.282 $\pm$ 0.107 & 0.0215 \\
 & 250 & 0.0764 $\pm$ 0.0019 & 0.2322 $\pm$ 0.0083 & 1.38 $\pm$ 0.13 & 0.0215 \\
 \hline
 \lno{} & 4.5 & 0.0629 $\pm$ 0.0022 & 0.292 $\pm$ 0.014 & 1.25 $\pm$ 0.13 & 0.0272 \\ 
 & 100 & 0.0796 $\pm$ 0.0024 & 0.279 $\pm$ 0.012 & 1.054 $\pm$ 0.087 & 0.0272 \\
 & 250 & 0.0848 $\pm$ 0.0015 & 0.2278 $\pm$ 0.0058 & 1.67 $\pm$ 0.13 & 0.0272 \\
 & 320 & 0.0883 $\pm$ 0.0026 & 0.2282 $\pm$ 0.0093 & 1.68 $\pm$ 0.21 & 0.0272 \\
 [1ex] 
 \hline
\end{tabular}
\caption{Parameters obtained from a stretched exponential fit to the ZF asymmetry histograms as described in the main text.}
\label{FitParameters}
\end{table}

Table \ref{FitParameters} contains the parameters obtained from a stretched exponential (plus background) fit to the zero field histograms listed in Table \ref{runlogzf}. 

Table \ref{Escanfit} shows the resulting fit parameters from a minimal model of discontinuous asymmetry changes within the multilayer samples. $A_{RNO}$ and $A_{STO}$ are determined from the sharp interface model where the asymmetry changes only at the chemical interfaces within the multilayer thus the only free parameters are the ``bulk''-like asymmetries of the nickelate layer and the \sto{} respectively. This model generates the dashed lines plotted in Figure 4 of the main text.

The second type of model considered has an additional free parameter, here called $d_l$, that indicates how far into the substrate the ``bulk'' asymmetry, $A'_{RNO}$, of the nickelate layer may be found before the asymmetry reverts discontinuously to the ``bulk'' value for \sto{} $A'_{STO}$.

The limited range of these stray fields make ferromagnetism an unlikely scenario, as discussed in the main text.

Note that the experimental asymmetries plotted in Figure 4 have been corrected from the initial output of the decaying cosine fit by subtracting the asymmetry deriving from the muons lost to backscattering at low energies. The function subtracted as correction is $A_c=0.22e^{-E/1.26} + 0.0019$, where $E$ is in keV.

\begin{table}[]
\centering
\begin{tabular}{|c || c | c || c | c | c |} 
 \hline
 Sample & $A_{RNO}$ & $A_{STO}$ & $A'_{RNO}$ & $A'_{STO}$ & $d_l$ (nm)\\ [0.5ex] 
 \hline
 \nsno{} & 0.019 $\pm$ 0.005 & 0.048 $\pm$ 0.002 & 0.0295 $\pm$ 0.0005 & 0.0495 $\pm$ 0.0005 & 4.5 $\pm$ 1.5\\ 
 \psno{} & 0.037 $\pm$ 0.004 & 0.036 $\pm$ 0.001 & 0.0357 $\pm$ 0.0010 & 0.0404 $\pm$ 0.0004 & 8.2 $\pm$ 0.7\\
 \lsno{} & 0.034 $\pm$ 0.005 & 0.048 $\pm$ 0.001 & 0.0415 $\pm$ 0.0013 & 0.0516 $\pm$ 0.0005 & 5.2 $\pm$ 0.7\\
 \lno{} & 0.022 $\pm$ 0.003 & 0.042 $\pm$ 0.001 & 0.0367 $\pm$ 0.0007 & 0.0454 $\pm$ 0.0005 & 6.5 $\pm$ 0.2\\
 \hline
\end{tabular}
\caption{Bulk asymmetries obtained by fitting a minimal model to the energy-dependent wTF asymmetries.}
\label{Escanfit}
\end{table}

\FloatBarrier

\bibliography{PAPERS-MuSR}